\documentclass[12pt,preprint]{aastex}

\newcommand{\logg} {\log g}
\newcommand{\halpha} {H$\alpha$}

\newcommand{\Te} {T_{\rm eff}}

\newcommand{\mv} {$M_V$}
\newcommand{\mb} {$M_B$}

\newcommand{\msun} {$M_\odot$}

\newcommand{\vtan} {v_{\rm tan}}
\newcommand\kms{km~s$^{-1}$}
\newcommand\gta{\lower 0.5ex\hbox{$\buildrel > \over \sim\ $}} 
\newcommand\lta{\lower 0.5ex\hbox{$\buildrel < \over \sim\ $}} 

\shortauthors{Bergeron}
\shorttitle{Halo White Dwarfs}
\begin{document}

\title{A Critical Examination of Halo White Dwarf Candidates}

\author{P. Bergeron}
\affil{D\'epartement de Physique, Universit\'e de Montr\'eal, C.P.~6128, 
Succ.~Centre-Ville, 
Montr\'eal, Qu\'ebec, Canada, H3C 3J7.}
\email{bergeron@astro.umontreal.ca}

\begin{abstract}

A detailed analysis of halo white dwarf candidates is presented, which
is based on model atmosphere fits to observed energy distributions
built from photoelectric or photographic magnitudes. Most of the
candidates identified in reduced proper motion diagrams are shown to
be too warm ($\Te>5000$~K) and most likely too young to be members of
the galactic halo, while the tangential velocities of the cooler and
thus older white dwarfs are shown to be entirely consistent with the
disk population. The results suggest that some white dwarf stars born
in the young disk may have high velocities with respect to the local
standard of rest. Such objects could represent the remnants of donor
stars from close mass-transfer binaries that produced type Ia
supernovae via the single degenerate channel, or other scenarios
suggested in the literature. Ongoing surveys that rely solely on
reduced proper motion diagrams are likely to identify more of these
high velocity young degenerates, rather than to unveil the old white
dwarf population of the galactic halo. The importance of infrared
photometry for studying extremely cool white dwarfs is also
emphasized.

\end{abstract}


\section{Introduction}

The first white dwarf stars to have formed in the galactic halo are
expected to be extremely old ($\tau>14$ Gyr), cool ($\Te < 4000$~K),
and have low luminosities ($L/L_\odot < 10^{-5}$). They should also
have large proper motions as a result of their high velocities with
respect to the local standard of rest. The latter defines the frame of
reference near the sun that rotates about the center of the Galaxy
with a velocity of $\sim 220$ \kms. Moreover, if these white dwarfs
have hydrogen-rich atmospheres, the collision-induced absorptions by
molecular hydrogen would make these objects appear bluish in
color-magnitude diagrams
\citep[see, e.g.,][]{hansen98}. These combined characteristics have 
led several investigators to look for this halo population of large
proper motion white dwarfs by using digitized photographic plates
taken at various epochs, and in particular by studying the location of
individual stars in the so-called reduced proper motion diagrams. The
identification of this halo population will help to estimate the
contribution of white dwarf stars to the mass budget of the galactic
halo, and determine whether they could account for the reported
microlensing events.

The direct detection of galactic halo dark matter has been claimed by
\citet[][hereafter OHDHS]{opp01} who identified 38 cool halo white dwarf candidates in the
SuperCOSMOS Sky Survey, with an inferred space density that could
account for 2\% of the halo dark matter. This conclusion, however, has
been challenged repeatedly since then by several authors
\citep[see][and references therein]{torres02}. While some studies
have reinterpretated the white dwarf kinematics \citep{reid01}, others
deal with a revaluation of the distances and tangential velocities
\citep{torres02} or age distribution \citep{hansen01}. 

In this paper, we take another look at halo white dwarf candidates
reported in the literature. Our approach differs from previous
investigations, however, in that we go back to the original published
magnitudes -- photoelectric or photographic, and attempt to extract
the fundamental stellar parameters using model atmosphere fitting
techniques that were developed to study the cool white dwarfs in the
galactic disk. We first analyze halo white dwarf candidates with
published trigonometric parallax measurements for which stellar masses
and cooling ages can be determined. Then we reexamine the sample of
white dwarfs reported by OHDHS using a similar approach.

\section{White Dwarfs with Trigonometric Parallax Measurements}\label{sbsc:trig}

\citet{ldm89} were the first to identify white dwarfs in the solar
neighborhood that could be interpreted as interlopers from the halo
population of the Galaxy. Their Figure 2 shows the distribution of
tangential velocities ($\vtan$) with \mv\ for the sample used by
\citet{ldm88} to measure the cool white dwarf luminosity function,
augmented by white dwarfs in the LHS Eight Tenths Sample brighter than
$M_V=+13$. Six white dwarfs in this diagram had tangential velocities
in excess of $\vtan=250$ \kms\ that could be associated with the thick
disk or halo populations. We reproduce in Figure \ref{fg:f1} a similar
result but for the white dwarf sample of \citet[][hereafter
BLR]{blr01} composed of cool degenerates with trigonometric parallax
and proper motion measurements, from which distances and thus
tangential velocities can be determined. Note that \mb\ is used here
instead of \mv\ in order to overplot the results of OHDHS
discussed below.  Five of the six halo white dwarf candidates from
\citet{ldm89} --- LHS 56, LHS 147, LHS 282, LHS 291, and LHS 524 ---
are labeled in Figure
\ref{fg:f1}; the missing object (LHS 2984) is below the range of the
plot. We also marked three additional objects with large tangential
velocities (G126$-$25, G138$-$56, and L845$-$70). The results displayed
here indicate that most of these halo white dwarf candidates have
parallax measurements with large uncertainties, however, which
translate into large errors in both \mb\ and $\vtan$. Only LHS 56 and
LHS 542 have errors smaller than 30\%, while three objects have errors
even larger than the measurement itself. The remaining white dwarfs in
Figure \ref{fg:f1} have estimated tangential velocities below $\sim
175$ \kms, consistent with the disk population.

Also shown in Figure \ref{fg:f1} are the results obtained for the 38
white dwarfs taken from Table 1 of OHDHS. The \mb\ values are derived
from their photographic color-magnitude calibration, $M_{B_{\rm
J}}=12.73+2.58\,(B_{\rm J}-R_{\rm 59F})$, and from the color equation
of \citet{blair82}. As can be seen, a significant fraction of this
sample stands out from the disk white dwarfs studied by BLR. This is
actually the main argument used by OHDHS to interpret these high
velocity stars as halo members. These results are discussed further in
\S~\ref{sbsc:opp} below.

We now proceed to analyze the optical $BVRI$ and infrared $JHK$
photometry (CCD) for the eight halo white dwarf candidates identified in Figure
\ref{fg:f1} (only three
objects have infrared photometry available) using the model
atmospheres and fitting technique described at length in
\citet[][hereafter BRL]{brl97}. Briefly, the optical and infrared magnitudes are first
converted into average fluxes, which are then compared with those
obtained from the model atmospheres --- properly averaged over the
filter bandpasses --- using a nonlinear least-squares method. Only
$\Te$ and the solid angle $\pi(R/D)^2$ are considered free
parameters. The distance $D$ is obtained from the trigonometric
parallax measurement, and the stellar radius $R$ is converted into
mass using the C/O-core evolutionary models described in \citet{blr01}
with thin or thick hydrogen layers, which are based on the
calculations of \citet{fon01}. Note that a value of $\logg=8.0$ was
assumed for four of these stars --- LHS 282, G138$-$56, L845$-$70, and
G126$-$25 --- since their trigonometric parallax measurements are too
uncertain to yield any meaningful value of the stellar radius and
mass. Sample fits for the three white dwarfs with complete $BVRI$ and
$JHK$ energy distributions are displayed in Figure \ref{fg:f2}, where
both the monochromatic and average model fluxes are shown, although
only the latter are used in the fitting procedure. High
signal-to-noise spectroscopic observations at \halpha\ (not shown
here) confirm our solutions (see the discussion in BLR). We note
already that the stars in Figure \ref{fg:f2} have temperatures and
masses that are quite similar to the $\sim 150$ disk white dwarfs
analyzed by BLR.

To estimate the age of these halo white dwarf candidates, we plot in
the bottom panel of Figure \ref{fg:f3} their location in a mass versus
$\Te$ diagram together with the isochrones from our C/O-core white
dwarf cooling models with thick hydrogen layers; thin hydrogen layer
models would yield ages that are $\sim 1.5$ Gyr {\it younger}. Also
shown are the corresponding isochrones but with the main sequence
lifetime added to the white dwarf cooling age; here we simply assume
\citep{lrb98} $t_{\rm MS}=10(M_{\rm MS}/M_\odot)^{-2.5}$ Gyr and
$M_{\rm MS}/M_\odot=8\ln[(M_{\rm WD}/M_\odot)/0.4]$. It is clear that
none of these white dwarfs are old enough to be members of the
galactic halo. On the contrary, they all lie comfortably within the
bulk of other (young) disk white dwarfs taken from BLR, also displayed
in the figure. Of course, several of the halo candidates selected from
Figure \ref{fg:f1} would have not been considered in the first place
if the parallax measurement had been more accurate. But even LHS 56
($\Te=7270$~K) and LHS 542 ($\Te=4720$~K, shown in Fig.~\ref{fg:f2}),
which have the best trigonometric parallax measurements, are about 2.0
and 8.5 Gyr old, respectively. This is considerably younger than what
is expected from a stellar halo population, typically older than $\sim
14$ Gyr.

These results, taken at face value, suggest that there exists a
population of young white dwarfs, most likely formed in the thin disk,
with large tangential velocities typical of the halo population
($\vtan > 250$ \kms). The possible nature of these high velocity white
dwarfs is discussed below in \S~\ref{sbsc:disc}. It also appears from
our results that there are no white dwarfs with measured trigonometric
parallaxes that are members of the galactic halo. Hence the luminosity
function of halo white dwarfs determined by \citet{ldm89} and used by
\citet{chabrier96} to estimate the white dwarf contribution to the
halo missing mass is to be considered premature. With these results in
mind, we now turn to the OHDHS sample.

\section{White Dwarfs in the Oppenheimer et al.~sample}\label{sbsc:opp}

\subsection{Reduced Proper Motion Diagram}

We first begin by studying the reduced proper motion diagram, which is
commonly used to identify halo white dwarf candidates in surveys based
on digitized photographic plates \citep[see, e.g.,][OHDHS]{knox99,
hambly01}. The reduced proper motion is defined as $H_{\rm R}=R_{\rm
59F}+5\log\mu+5$, where $R_{\rm 59F}$ is the photographic magnitude,
and $\mu$ is the proper motion measured in arc seconds per year. As
explained by \citet{knox99}, $H_{\rm R}$ is an intrinsic property of
the star that gives some estimate of its absolute magnitude. Stars
that are relatively blue and with large values of $H_{\rm R}$ in this
diagram are viewed as good halo white dwarf candidates since old, and
thus cool, white dwarfs have low luminosities and turn blue below
$\sim 3500$~K \citep[see][for instance]{hambly01}. 

We show in Figure \ref{fg:f4} the reduced proper motion diagram for
the combined sample of disk white dwarfs from BRL and BLR (note that
trigonometric parallaxes are not required to construct this diagram)
as well as the OHDHS sample. Since the OHDHS analysis relies on the
photographic $B_{\rm J}-R_{\rm 59F}$ color index, the $B$ and $R$
magnitudes from BRL and BLR had to be converted into photographic
magnitudes using the equations of \citet{blair82} and
\citet{bessell86}. There are two white dwarfs, LHS 147 and LHS 542,
that are in common between the BRL/BLR and the OHDHS samples, and
their location in Figure \ref{fg:f4} are in excellent agreement.
Five of the six halo white dwarf candidates identified by
\citet{ldm89} and analyzed in the previous section are labeled 
in Figure \ref{fg:f4}. Obviously, these objects clearly stand out from
the rest of the BRL and BLR samples of disk white dwarfs, but as shown
above, this is mostly due to their large proper motions and not to
their low luminosities. Their relatively blue color indices are
representative of white dwarfs with effective temperatures well above
4000 K, and not indicative of extremely cool degenerates. So what we
are seeing here is only a {\it vertical shift} with respect to the
disk sequence.

The 38 white dwarfs from the OHDHS sample are displayed
in Figure \ref{fg:f4} as well (the different symbols are explained in
\S~\ref{sbsc:disc}). We already note that about half of this sample 
overlaps completely with the disk population, and there is thus no
indication, at least from this diagram, that this subsample should be
associated with an older population of the Galaxy. Only the other half
of the sample actually exhibits the required characteristics to be
halo white dwarfs. It is interesting to note, however, that the
candidates from
\citet{ldm89} represent even more extreme cases than the OHDHS sample,
yet none are halo members according to our detailed model atmosphere
analysis.

The horizontal dotted line in Figure
\ref{fg:f4} delineates the region below which the Luyten catalogs
contain very few objects according to OHDHS, primarily because
Luyten's survey in the Southern Hemisphere did not attempt to find
objects as faint as those found by OHDHS. Interestingly enough, 10 out
of the 13 white dwarfs from the BRL/BLR sample found below this limit
are actually LHS stars. So with the exception of F351$-$50 and
WD0351$-$564, the two reddest objects at the bottom of Figure
\ref{fg:f4}, there is a considerable overlap in this diagram between
these samples, despite the fact that OHDHS searched for objects as
faint as $R_{\rm 59F}=19.8$.

\subsection{Color-Magnitude Diagrams}

In order to estimate the tangential velocity of the white dwarfs in
their sample, OHDHS relied on distances obtained from the photographic
magnitudes $B_{\rm J}$ and their color-magnitude relation ---
$M_{B_{\rm J}}=12.73+2.58\,(B_{\rm J}-R_{\rm 59F})$ --- derived from a
linear least-squares fit to the cool white dwarf sample of BLR. Figure
\ref{fg:f5} shows the $M_{B_{\rm J}}$ versus $(B_{\rm J}-R_{\rm 59F})$
color-magnitude diagram for the BLR and OHDHS samples, together with
the theoretical sequences obtained from model atmospheres with pure
hydrogen and pure helium compositions. These models are similar to
those described in BLR and references therein \citep[see
also][]{bl02}.

Of course, the OHDHS data all appear on the same line since their
absolute magnitudes are based on a linear color-magnitude
relation. Hence in their analysis, a given value of $(B_{\rm J}-R_{\rm
59F})$ yields a unique value of $M_{B_{\rm J}}$, which combined with
$B_{\rm J}$ gives the distance. The results displayed in Figure
\ref{fg:f5} indicate that their relation is indeed a pretty good
match to the cool white dwarf sample of BLR, with the exception
perhaps at the hot end of the sequence where a small but significant
departure occurs. This could lead to distances being underestimated in
this range of effective temperature. A similar conclusion was reached
by \citet{torres02} who compared in their Figure 1 the OHDHS
calibration with a Monte Carlo simulation of the disk white dwarf
population (see comment below, however).

The main problem with this linear color-magnitude relation is that
hydrogen-rich atmosphere white dwarfs do not become increasingly red
as they cool off, but instead turn blue, as a result of the extremely
strong collision-induced absorptions by molecular hydrogen
\citep{hansen98,saumon99,bl02}. The effective temperature below which
this phenomenon occurs depends of course on the color index
considered. For instance, for a 0.6 \msun\ white dwarf, the turnoff in
$(V-I)$ occurs at $\Te\sim 3500$~K, while for the $(B_{\rm J}-R_{\rm
59F})$ color index discussed here, it occurs near $\Te\sim 3000$~K. It
is thus clear that the color-magnitude relation for hydrogen-rich
atmosphere white dwarfs cannot be linear. The same conclusion applies
to helium-rich atmosphere white dwarfs that contain only small traces
of hydrogen, since such stars have energy distributions that are
characterized by very strong infrared absorptions by molecular
hydrogen resulting from collisions with neutral helium
\citep[][]{bl02}. Hence for a given value of $(B_{\rm J}-R_{\rm
59F})$, or in general any color index in the visible region, there
exist {\it two} possible values of the absolute magnitude for white
dwarfs with atmospheres containing hydrogen.

Perhaps the best example is LHS 1402 whose spectrum is shown in
Figure 2b of OHDHS, and which according to the authors is a cooler
analog of LHS 3250 and SDSS 1337$+$00, re-analyzed recently by
\citet{bl02}. Its measured value of $(B_{\rm J}-R_{\rm 59F})=0.46$ 
implies an absolute magnitude of $M_{B_{\rm J}}=13.92$, or a distance
of 76 pc ($B_{\rm J}=18.32$), i.e.~the value given in Table 1 of
OHDHS. However, this would also imply according to Figure
\ref{fg:f5} that LHS 1402 has $\Te\sim7000$~K, a temperature which can
be completely ruled out according to the observed spectrum. If one
assumes instead that LHS 1402 resides passed the turnoff on the pure
hydrogen sequence shown in Figure \ref{fg:f5}, one derives instead
$M_{B_{\rm J}}=19.97$ (for $M=0.6$ \msun), or a distance of only 4.7
pc! LHS 1402 may thus be among the closest white dwarfs to Earth. Of
course, a full model atmosphere analysis of this object is required
before reaching a firm distance estimate. The point remains that the
linear color-magnitude relation used by OHDHS neglects the fact that
there could be a substantial fraction of extremely cool white dwarfs
in their sample, and that the distances to these objects may be largely
overestimated. If this were true, the space density of halo white
dwarfs derived by OHDHS could even be higher than the value they inferred.

\subsection{Fits to the Energy Distributions}

It has been argued by BRL and BLR that the best way to determine the
fundamental parameters of cool white dwarf stars --- effective
temperature, mass, and atmospheric composition --- is to use the
entire energy distribution rather than to rely on the limited
information contained in color-color or color-magnitude diagrams. This
is particularly true if the $B$ filter is used, since for
hydrogen-rich atmosphere white dwarfs below $\Te\sim 5000$~K, there
seems to be a missing source opacity in the utlraviolet unaccounted
for in the model atmosphere calculations, resulting in an excess of
flux in this particular region of the spectrum \citep[see
BRL,][]{ber01}. This is the reason why in Figure \ref{fg:f5}, there
appears to be a deficiency of $\sim 0.6$ \msun\ hydrogen-atmosphere
white dwarfs below 5000 K, where all stars seem to follow instead the
helium-atmosphere sequence. As shown in Figure \ref{fg:f6}, this
discrepancy disappears if we consider a similar color-magnitude
diagram for the BLR sample, but this time using $M_R$ versus $R-I$ in
order to stay away from the ultraviolet regions. The location of
the hydrogen-rich and helium-rich atmosphere white dwarfs in this
diagram is consistent with the predicted colors.

It is worth mentioning here that the arguments presented by
\citet{torres02} with respect to the color-magnitude calibration used
by OHDHS, and in particular with the distances to the cool white
dwarfs being overestimated, are not completely accurate. The
calibration used by OHDHS has been determined empirically, and it is
based on a least-squares fit to the white dwarf observations from BLR,
as discussed above. In contrast, the Monte-Carlo simulations shown in
Figure 1 of Torres et al. are based on {\it theoretical colors} taken
from the cooling models of \citet{salaris00}. Hence the magnitude
discrepancy for the cool white dwarfs observed in their Figure 1 is
simply the result of the missing source of opacity in the hydrogen
model atmospheres from which the colors were calculated. Actually, the
color-magnitude calibration of OHDHS yields more accurate distance
estimates than the models, provided that the white dwarfs are hotter
than $\sim 3500$~K. Similarly, the conclusions of Torres et
al. regarding the objects lying beyond the turn-off in Figure
\ref{fg:f5} as being necessarily helium-atmosphere white dwarfs
is also misleading since these stars could as well have hydrogen
atmospheres if one considers the more appropriate $M_R$ versus $R-I$
diagram displayed in Figure \ref{fg:f6}. Such ambiguities can be
easily avoided by taking into account the complete energy
distributions.

Ideally, energy distributions should be built from optical $BVRI$ and
infrared $JHK$ photometric measurements, but such measurements are not
available for the moment for the OHDHS sample. Nevertheless, we can
get some insight into the nature of these white dwarfs by using the
photographic $BRI$ magnitudes published by OHDHS. We thus proceed with
fitting these white dwarfs by first transforming the photographic
magnitudes into standard $BRI$ magnitudes using the equations of
\citet{blair82} and \citet{bessell86}, as well as the empirical 
relation $(B-V)=0.097+0.612\,(R-I)+2.136\,(R-I)^2$ derived from the BRL
and BLR observations. These magnitudes are then converted into
broadband fluxes and compared with the predictions of our model
atmospheres using the fitting technique described in
\S~\ref{sbsc:trig}. Since there is no trigonometric parallax
measurement for these stars, a value of $\logg=8.0$ is assumed for all
objects.

With the lack of infrared measurements, it may be difficult to
determine whether a cool white dwarf has a hydrogen- or helium-rich
atmosphere. For stars above $\Te\sim 5000$~K, however, BRL have shown
that the presence of \halpha\ implies in general that the white dwarf
has a hydrogen-rich atmosphere, while the absence of this feature
implies a helium-rich atmosphere. We have thus used the spectroscopic
information available in Table 1 of OHDHS to establish the atmospheric
composition of all stars above $\Te=5000$~K. This was possible for 20
out of 38 white dwarfs in the sample. Typical fits are displayed in
Figure \ref{fg:f7} for hydrogen and helium atmosphere white
dwarfs. Predicted broadband fluxes are shown at $BVRI$ as well as
$JHK$. For stars below 5000 K, it is not possible to determine the
atmospheric composition unambiguously, as shown for the two objects at
the bottom of Figure \ref{fg:f7}. Clearly, infrared measurements would
help to discriminate between the two solutions. The atmospheric
composition can nevertheless be inferred for two cool white dwarfs in
the sample: LHS 542, which has been analyzed by BLR, has a complete
energy distribution consistent with a helium-rich atmosphere (see
Fig.~\ref{fg:f2}). Also, the spectrum of LHS 1402 displayed in Figure
2 of OHDHS reveals a very strong near infrared flux deficiency,
suggesting a hydrogen-rich composition. We are thus left with 16 out
of 38 stars with no information about the atmospheric composition.

The problem becomes even more complex when one considers the results
shown in Figure \ref{fg:f5}, which reveal that for some cool stars,
three solutions are possible: a cool helium atmosphere, a cool ($\Te\
\lta 3000$~K) or a hot ($\Te\ \gta 3000$~K) hydrogen atmosphere. In
the absence of infrared photometric measurements, it is difficult to
distinguish between these three solutions. In addition, because of the
missing source of UV opacity in the hydrogen model atmospheres (see
discussion above), it may even be dangerous to rely on the $B$
magnitudes for these cool white dwarfs. We have thus followed the
prescription of BLR and omitted the $B$ filter when fitting the
coolest white dwarfs in the sample with pure hydrogen models. Our
fitting procedure is illustrated in Figure \ref{fg:f8} for the reddest
objects in the color-magnitude diagram shown in Figure
\ref{fg:f5}. F351$-$50 and WD0205$-$053 are so red that a unique
hydrogen solution is found, which corresponds to the hydrogen model
that produces the reddest possible value of $(R-I)$; the bad resulting
fit suggests that these stars probably have helium-rich atmospheres
instead. WD0351$-$564 and WD0345$-$362 on the other hand have two
possible hydrogen solutions that differ only by the quality of the fit
at $B$. The large discrepancy observed at $B$ for the warm hydrogen
solutions suggest that the cool solutions are probably more
appropriate. Again, it is obvious that infrared photometric
observations would help to constrain the solutions better. For the time
being, however, we will retain what appears to be the ``best''
hydrogen solution, as well as the helium solution, for all cool white
dwarfs in the sample.

To summarize our results, all white dwarfs with $\Te>5000$~K have
effective temperatures and atmospheric compositions that are well
constrained by our fitting technique, while cooler stars possess at
least two possible solutions that depend on the assumed atmospheric
composition, but both solutions are nevertheless below 5000~K.

The accuracy of our fitting procedure using the limited information
contained in the energy distributions based solely on $BRI$ can be
assessed by comparing the effective temperatures of the two white
dwarfs in common between the OHDHS and the BLR samples for which we
have complete energy distributions. For LHS 147, we obtain
$\Te=7310$~K from $BRI$ alone, which can be compared with $\Te=7600$~K
when the complete energy distribution is used (see Fig.~\ref{fg:f2}),
whereas for LHS 542 we derive $\Te=4780$~K versus 4720~K
under similar conditions. This comparison indicates that the
temperature scale obtained here is quite reliable for the purpose of
our discussion.

\section{Results and Discussion}\label{sbsc:disc}

The effective temperature distribution of the halo white dwarf
candidates in the OHDHS sample is displayed as a histogram in the top
panel of Figure \ref{fg:f3}, with the number of stars in each 1000~K
bin shown on the left hand side of the figure. For the cooler stars,
we have taken the average of the hydrogen and helium atmosphere
solutions; there is also one star, LP 586$-$51, with a temperature of
$\sim 14,000$~K outside the range of the plot. About 75\% of the stars
in the OHDHS sample have effective temperatures above 4000~K, and
unless their masses are very close to 0.45 \msun, they are obviously
too young to be associated with the halo population. Even though the
halo membership of these stars cannot be completely ruled out until
precise trigonometric parallax measurements are secured, it seems more
reasonable, in the light of the results obtained for the halo white
dwarf candidates identified by \citet{ldm89}, also shown in Figure
\ref{fg:f3}, to associate these OHDHS stars with the same population
of young, high velocity white dwarfs.

We note also in Figure \ref{fg:f3} that all seven stars in the
3000-4000~K bin have effective temperatures taken as the average of
the hydrogen and helium atmosphere solutions below and above
$\Te=4000$~K, respectively. If infrared measurements confirm that
these stars have helium atmospheres, this particular bin could end up
empty! Also, the object in the coolest bin corresponds to LHS 1402,
the white dwarf with the strong infrared flux deficiency discussed
above. As demonstrated by \citet{bl02}, if this flux deficiency is the
result of collision-induced absorptions by molecular hydrogen due to
collisions with {\it neutral helium} in a helium-rich atmosphere,
rather than with other hydrogen molecules in a hydrogen-rich
atmosphere, the effective temperature derived here from pure hydrogen
models could be significantly underestimated. This would leave no
ultracool, and presumably old, white dwarfs in the OHDHS sample.

Probably the most puzzling result of our analysis is illustrated in
Figure \ref{fg:f4} where white dwarfs with $\Te>5000$~K are shown as
filled diamonds, while cooler stars are indicated by filled
circles. These results reveal that the hottest stars in the OHDHS
sample correspond to the objects that lie below the sequence defined
by the disk white dwarfs, with only a few cooler white dwarfs found in
the same region of the diagram. The object at $H_{\rm R}=21.3$ and
$(B_{\rm J}-R_{\rm 59F})=0.46$ is LHS 1402, the coolest white dwarf in
the sample; there is also LHS 542, labeled in the figure, which we
know is too young to be a halo white dwarf. With the additional
exception of F351$-$50 and WD0351$-$564 at the bottom of the diagram,
all white dwarfs with $\Te<5000$~K in the OHDHS sample overlap
comfortably with the disk white dwarfs analyzed by BRL and BLR. There
is thus no reason left to believe that even the cool stars in the
OHDHS sample belong to an old galactic population, either from
kinematics or age considerations. We are thus forced to conclude that
white dwarfs identified in such reduced proper motion diagrams, and
interpreted as halo candidates, should rather be associated with a
hotter and presumably younger population of the Galaxy. We note that
LHS 1402, probably the most promising halo candidate in the OHDHS
sample, does not even particularly stand out in this reduced proper
motion diagram.

It is interesting to mention in this context that \citet{hambly01} has
identified three halo white dwarf candidates from such a reduced
proper motion diagram (see his Figure 1). His most extreme case (point
indicated by a circle in his diagram) has a large reduced proper
motion and relatively blue color, which make this object a very good
halo candidate according to Hambly. However, from the published $BRI$
colors, we derive an effective temperature of $\sim 7200$~K for this
star using our fitting procedure, which suggests that this object
also belongs to the same population of young high velocity white
dwarfs found in the \citet{ldm89} and OHDHS samples.

Similar conclusions can be reached if we consider once again the
distribution of tangential velocities with \mb, shown in Figure
\ref{fg:f9}, but this time using our own distance estimates for the
calculations of $\vtan$. As can be seen, all white dwarfs in the OHDHS
sample with $M_B\ \lta 15$ have high tangential velocities
characteristic of the halo population. Yet, according to our $\Te$
determinations, these are probably too hot and too young to belong to
the halo population. For $M_B\ \gta 15$, the location of all white
dwarfs in the OHDHS sample is consistent with the disk population,
with the exception of LHS 542, F351$-$50, and WD0351$-$564. As
mentioned above, LHS 542 is certainly too young to be a halo
member. At the top of this diagram lies LHS 1402, with a tangential
velocity entirely consistent with the disk population. Hence, even if
this star stands out marginally in the reduced proper motion diagram
shown in Figure \ref{fg:f4}, its tangential velocity is too small to
be associated with the halo. We note that LHS 1402 could be made much
more luminous ($M_B\sim 16$) if it has a helium-dominated atmosphere
similar to those of LHS 3250 and SDSS 1337$+$00 \citep[see Fig.~8
of][]{bl02}. We are thus left with only two objects in the OHDHS
sample, F351$-$50 and WD0351$-$564, that have the expected properties
of halo white dwarfs, that is high tangential velocities and low
effective temperatures, and even those could represent the cooled down
version of the young high velocity white dwarf population unveiled
here.

To summarize our findings, the OHDHS sample seems to be composed of
two distinct white dwarf components. The first one -- about half of
the sample -- is composed of cool white dwarfs which clearly belong to
the disk population, as there are no indication that any of these
stars differ from the other disk white dwarfs analyzed by BRL and
BLR. The second component is composed of relatively warm white dwarfs
with peculiarly high tangential velocities. Unless these white dwarfs
all have masses close to 0.45 \msun, they are too young to be
associated with the halo population, or even with the thick disk. One
argument in favor of these stars having normal masses ($M\sim 0.6$
\msun) and being relatively young is provided by the analysis of
the high velocity LHS stars with measured parallaxes, for which mass
estimates indicate ages much lower than 10 Gyr (see Figures
\ref{fg:f2} and \ref{fg:f3}). If this interpretation is correct, they
were probably born in the disk of the Galaxy, and somehow were
accelerated by some mechanism.

One such possible mechanism has been explored quantitatively by
\citet{hansen02} who proposed that these high velocity white dwarfs are
the remnants of donor stars from close mass-transfer binaries that
produced type Ia supernovae via the single degenerate channel. As the
close binary orbit gets disrupted by the supernova explosion, the
donor star is simply released, and eventually evolves to the white
dwarf stage with its pre-supernova high orbital velocity. Hansen
argues that the local density of such high velocity remnants is
comparable to that determined by OHDHS for their sample. One
consequence of this proposed scenario is that none of these white
dwarfs could have a binary companion. This seems at least consistent
with our results for the high velocity LHS stars, as none of them
appear to be overluminous (i.e. they are consistent with being single
stars). Other alternative mechanisms have been proposed by
\citet{davies02} and \citet{koopmans02} by which stars can be ejected 
from the thin disk into the galactic halo with the required high
velocities.

Further progress in answering some of the issues raised in this paper
can be achieved with the help of high quality $BVRI$ and $JHK$
photometric observations for the coolest stars in the OHDHS sample.
Trigonometric parallax measurements would also be required to measure
the individual white dwarf masses and obtain more accurate estimates
of the stellar ages. Such an endeavor is currently underway.

\acknowledgements This work was supported in part by the NSERC Canada 
and by the Fund NATEQ (Qu\'ebec).

\clearpage

\figcaption[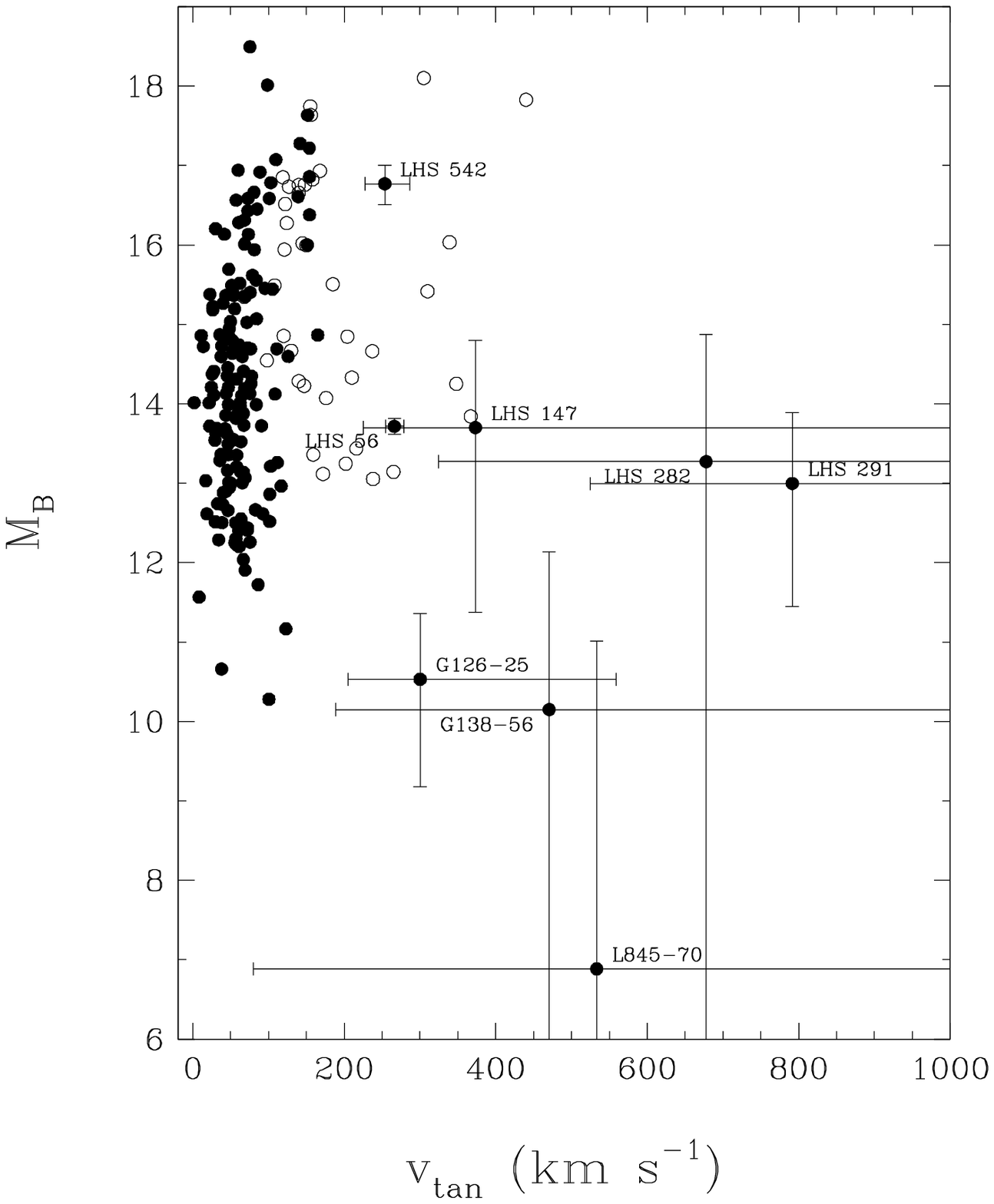] {Distribution of tangential velocities ($\vtan$) with the $B$ 
absolute magnitudes (\mb) for all cool ($\Te<12,000$~K) white dwarfs
with trigonometric parallax measurements ({\it filled circles}), and
for the halo white dwarf candidates taken from Table 1 of OHDHS ({\it
open circles}). The white dwarfs with known distances that have
tangential velocities consistent with a halo population are labeled
in the Figure, and the associated uncertainties are shown as well.
\label{fg:f1}}

\figcaption[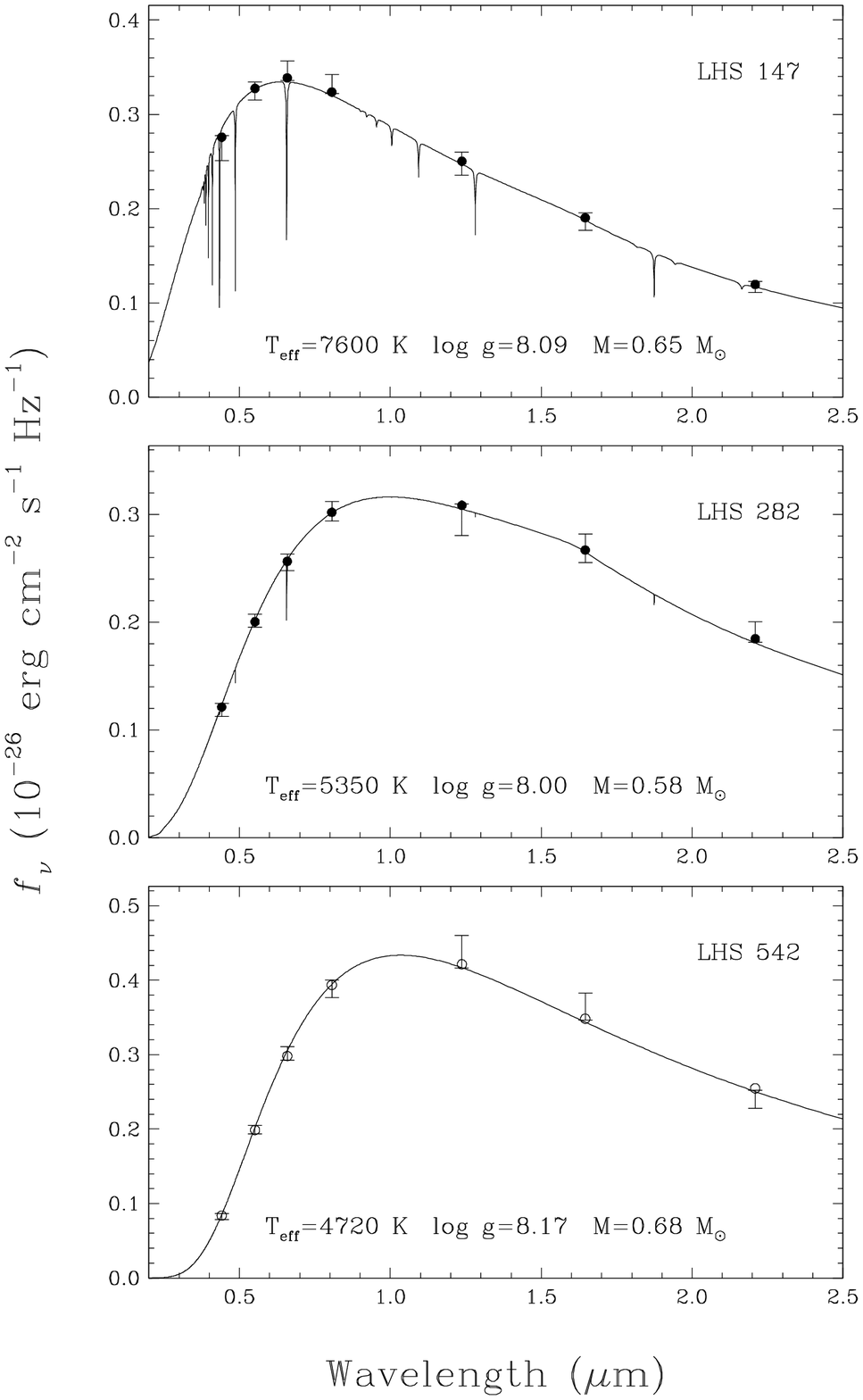] {Fits to the energy distributions of three halo white dwarf
candidates with pure hydrogen atmosphere ({\it filled circles}) or
pure helium atmosphere ({\it open circles}) models. The optical $BVRI$
and infrared $JHK$ photometric observations are shown by the error
bars. The solid lines correspond to the model monochromatic fluxes,
while the filled or open circles represent the average over the filter
bandpasses. For LHS 282, a value of $\logg=8.0$ was
assumed.\label{fg:f2}}

\figcaption[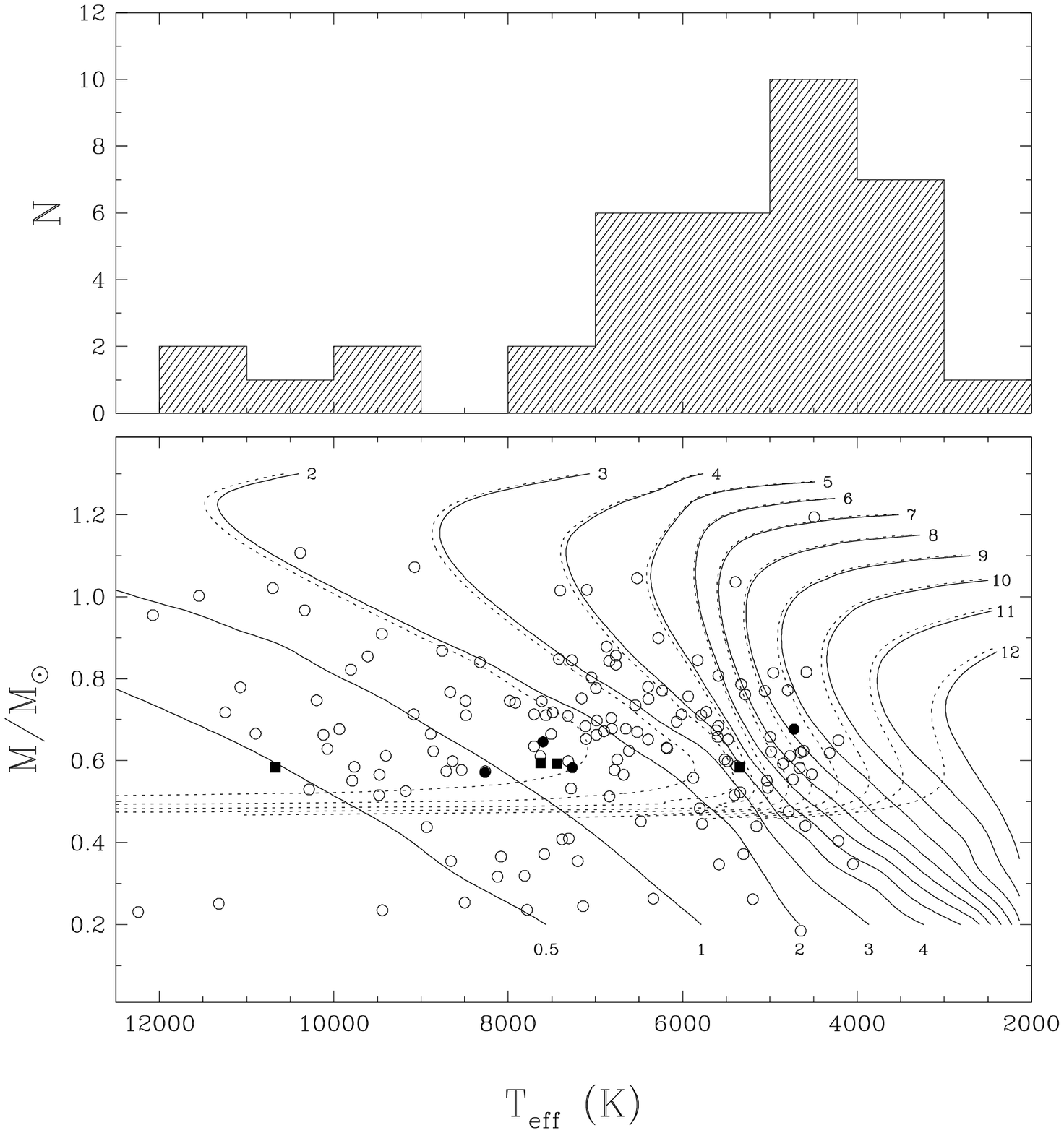] {Top panel: Masses of white dwarfs in the 
trigonometric parallax sample of BLR ({\it open circles}), and halo
white dwarf candidates ({\it filled symbols}; filled squares indicate
stars for which $\logg=8.0$ was assumed) as a function of effective
temperature, together with the isochrones from cooling sequences with
C/O-core, $q({\rm He})\equiv M_{\rm He}/M_{\star}=10^{-2}$, and
$q({\rm H})=10^{-4}$ ({\it solid lines}). The isochrones are labeled
in units of $10^9$ years. Also shown are the corresponding isochrones
with the main sequence lifetime taken into account ({\it dotted
lines}). Bottom panel: This histogram will be discussed in
\S~\ref{sbsc:disc}.\label{fg:f3}}

\figcaption[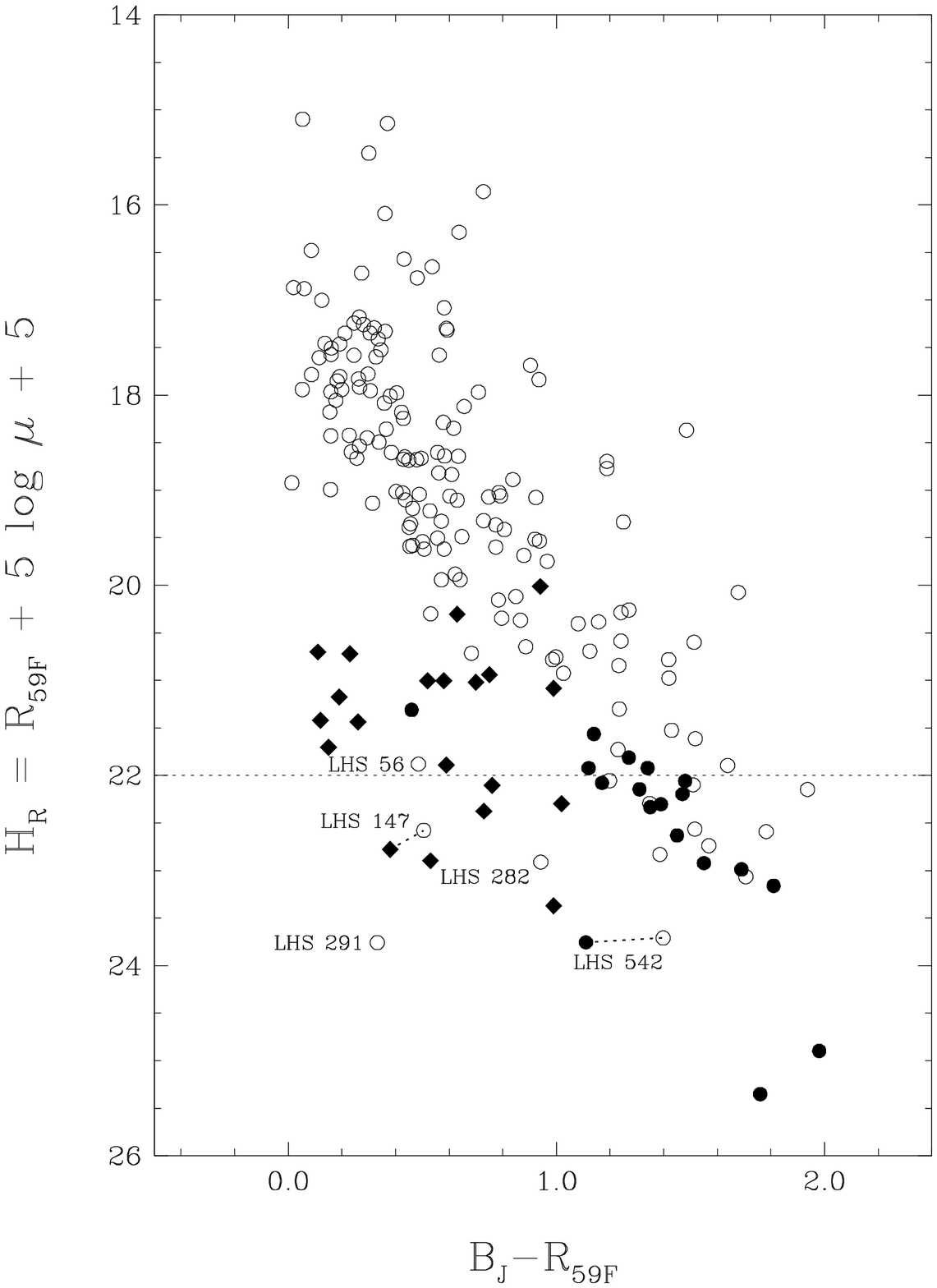] {Reduced proper motion diagram for the OHDHS
sample ({\it filled symbols}; the distinction between the filled
circles and filled diamonds is explained in
\S~\ref{sbsc:disc}) and the combined BRL and BLR samples ({\it open circles}).
The objects labeled correspond to the halo white dwarf candidates
identified by \citet{ldm89}, and analyzed in \S~\ref{sbsc:trig}. LHS
147 and LHS 542 are in common between the OHDHS and BRL samples, and
their individual measurements are connected by dotted lines. The
horizontal dotted line delineates the region below which the Luyten
catalogs contain very few objects according to OHDHS. The two reddest
white dwarfs at the bottom of the diagram are F351$-$50 and
WD0351$-$564.\label{fg:f4}}

\figcaption[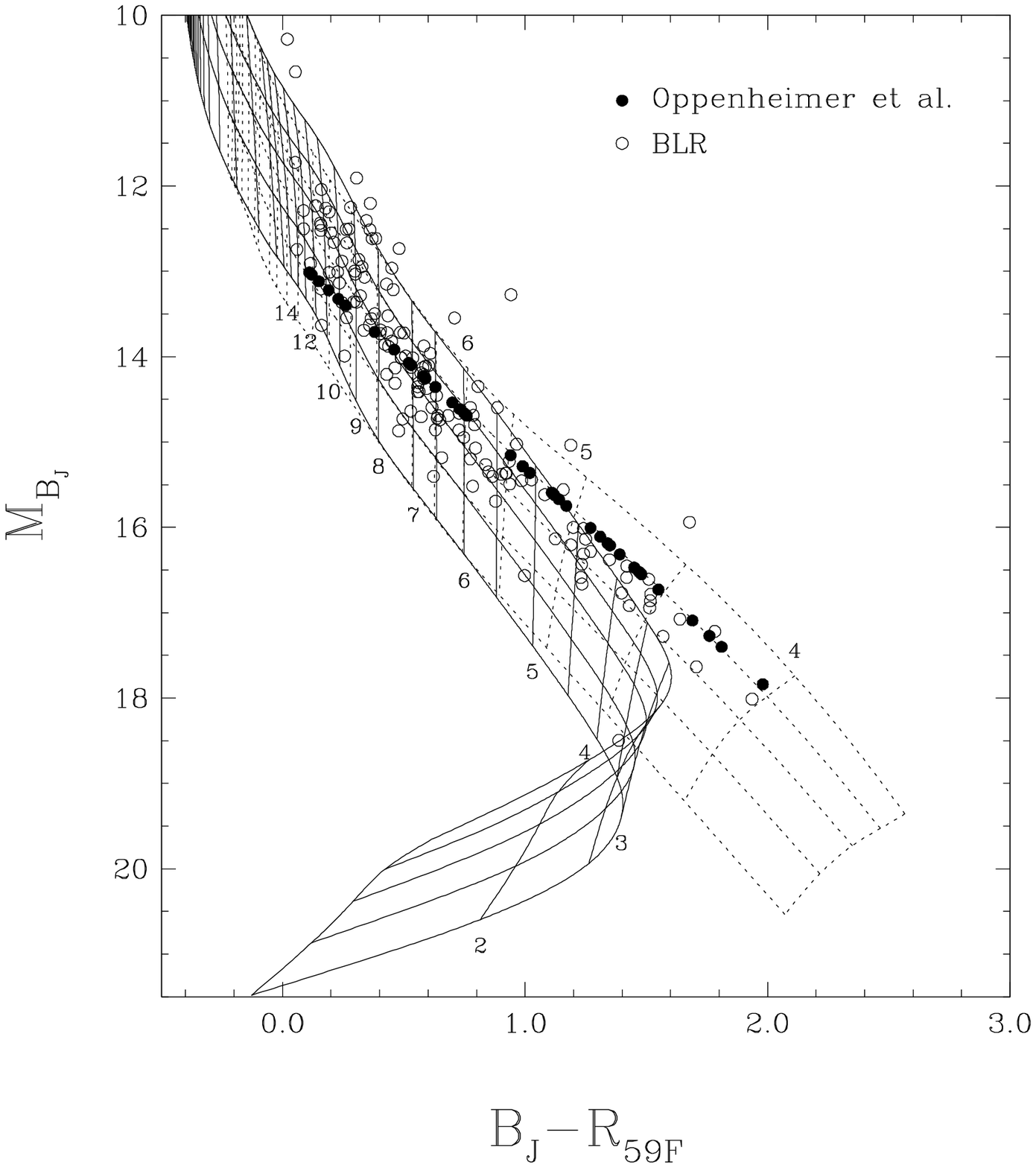] {$M_{B_{\rm J}}$ vs.~$(B_{\rm J}-R_{\rm 59F})$ 
color-magnitude diagram for the cool white dwarfs from BLR with
measured trigonometric parallaxes, together with the linear relation
used by OHDHS to estimate the absolute magnitudes for the stars in
their sample from the $(B_{\rm J}-R_{\rm 59F})$ photographic color
indices. Also shown are the theoretical colors for the pure hydrogen
({\it solid line}) and pure helium ({\it dotted line}) model
atmospheres with, from top to bottom, $M=0.4$, 0.6, 0.8, 1.0, and 1.2
\msun. Effective temperatures are indicated in units of $10^3$
K.\label{fg:f5}}

\figcaption[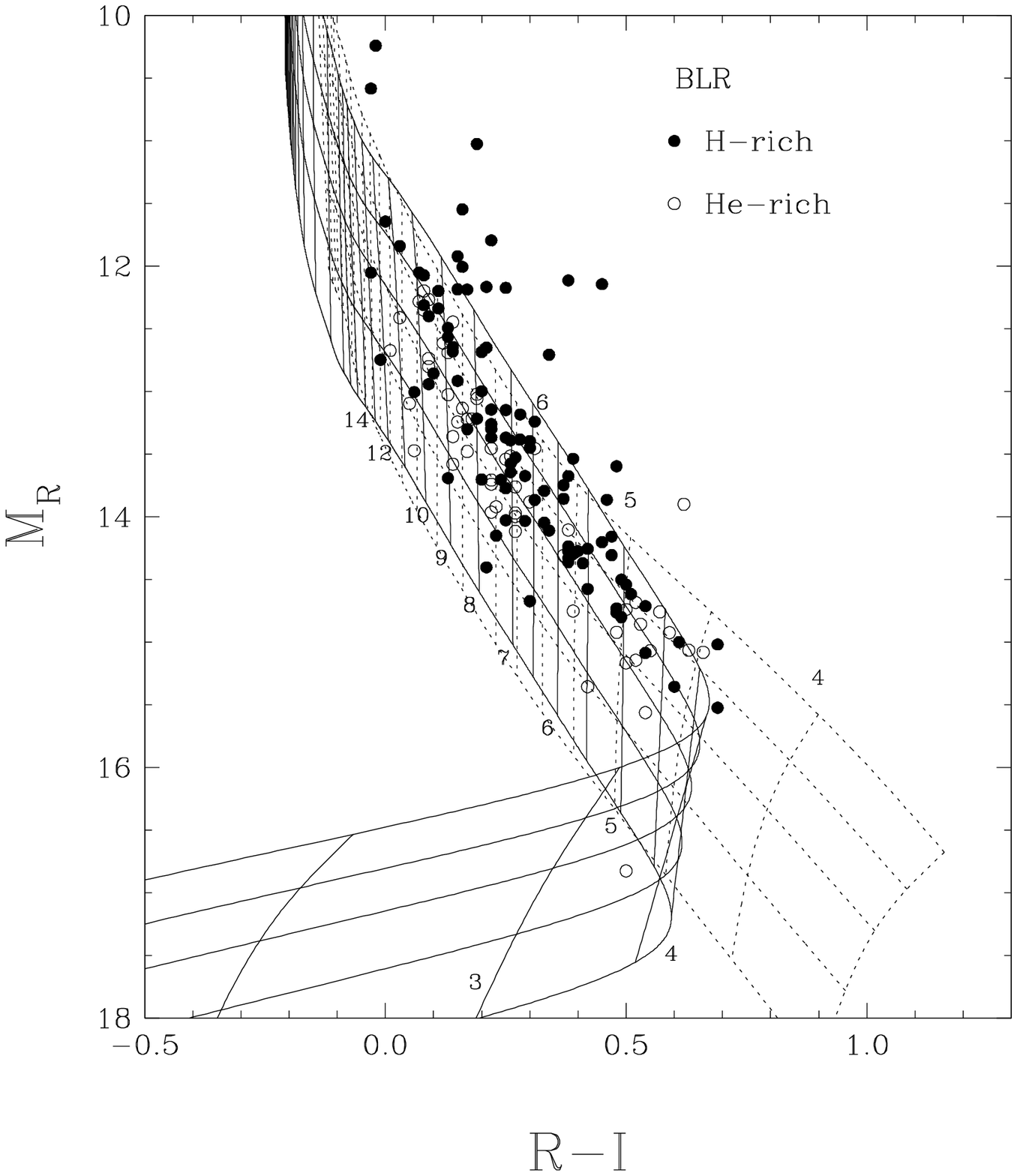] {$M_R$ vs.~$(R-I)$ color-magnitude diagram for 
the cool white dwarfs from BLR with hydrogen- ({\it solid dots}) and
helium-rich ({\it open circles}) atmospheres. Also shown are the
theoretical colors for the pure hydrogen ({\it solid line}) and pure
helium ({\it dotted line}) model atmospheres with, from top to bottom,
$M=0.4$, 0.6, 0.8, 1.0, and 1.2 \msun. Effective temperatures are
indicated in units of $10^3$ K.\label{fg:f6}}

\figcaption[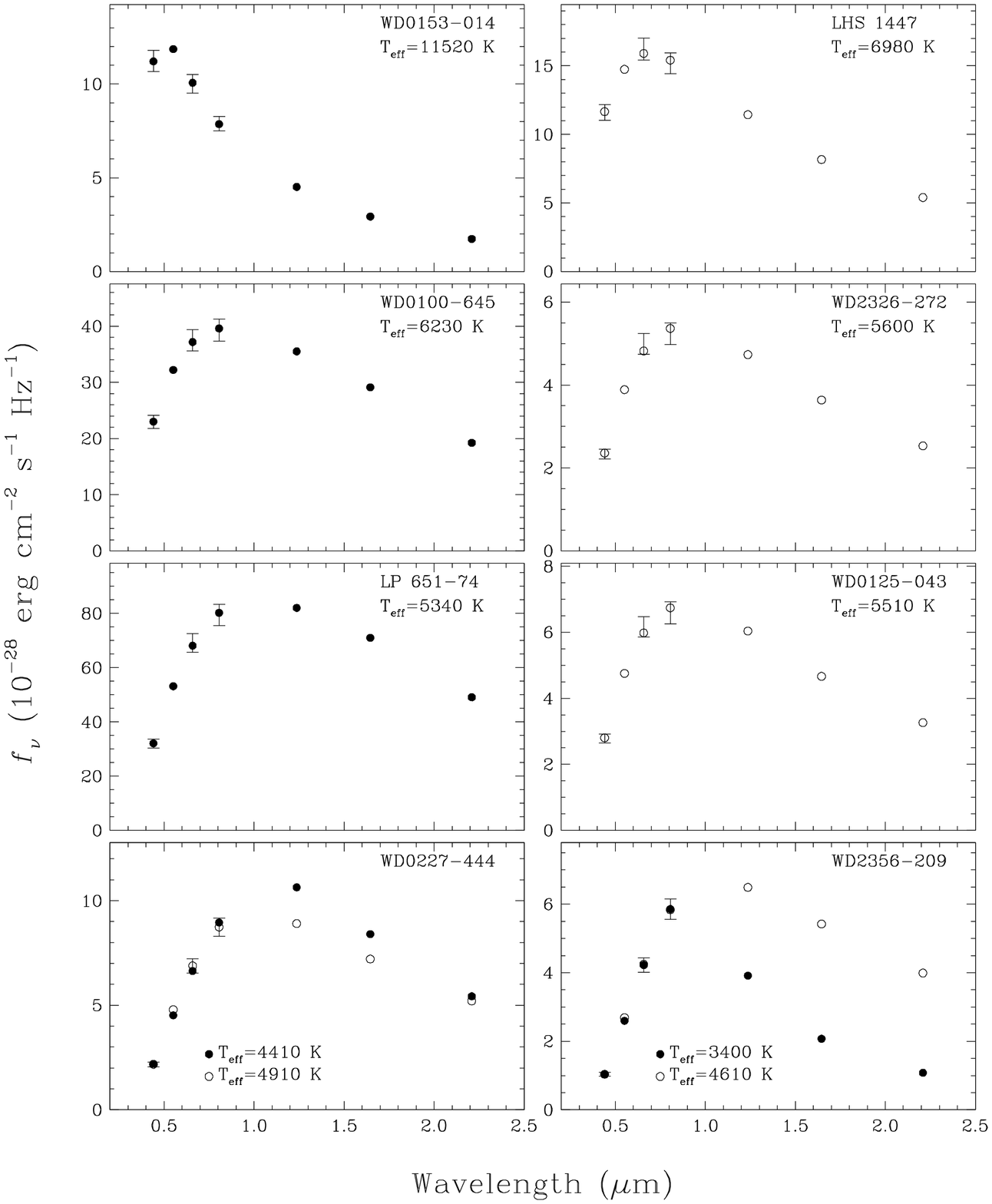] {Typical fits to the energy distributions of
halo white dwarf candidates from OHDHS with pure hydrogen atmosphere
({\it filled circles}) or pure helium atmosphere ({\it open circles})
models. The optical $BRI$ photometric observations are shown by the
error bars, while the filled or open circles represent the predicted
model broadband fluxes et $BVRIJHK$. A value of $\logg=8.0$ was assumed for all
objects. Two solutions are possible for the coolest white dwarfs shown
at the bottom.\label{fg:f7}}

\figcaption[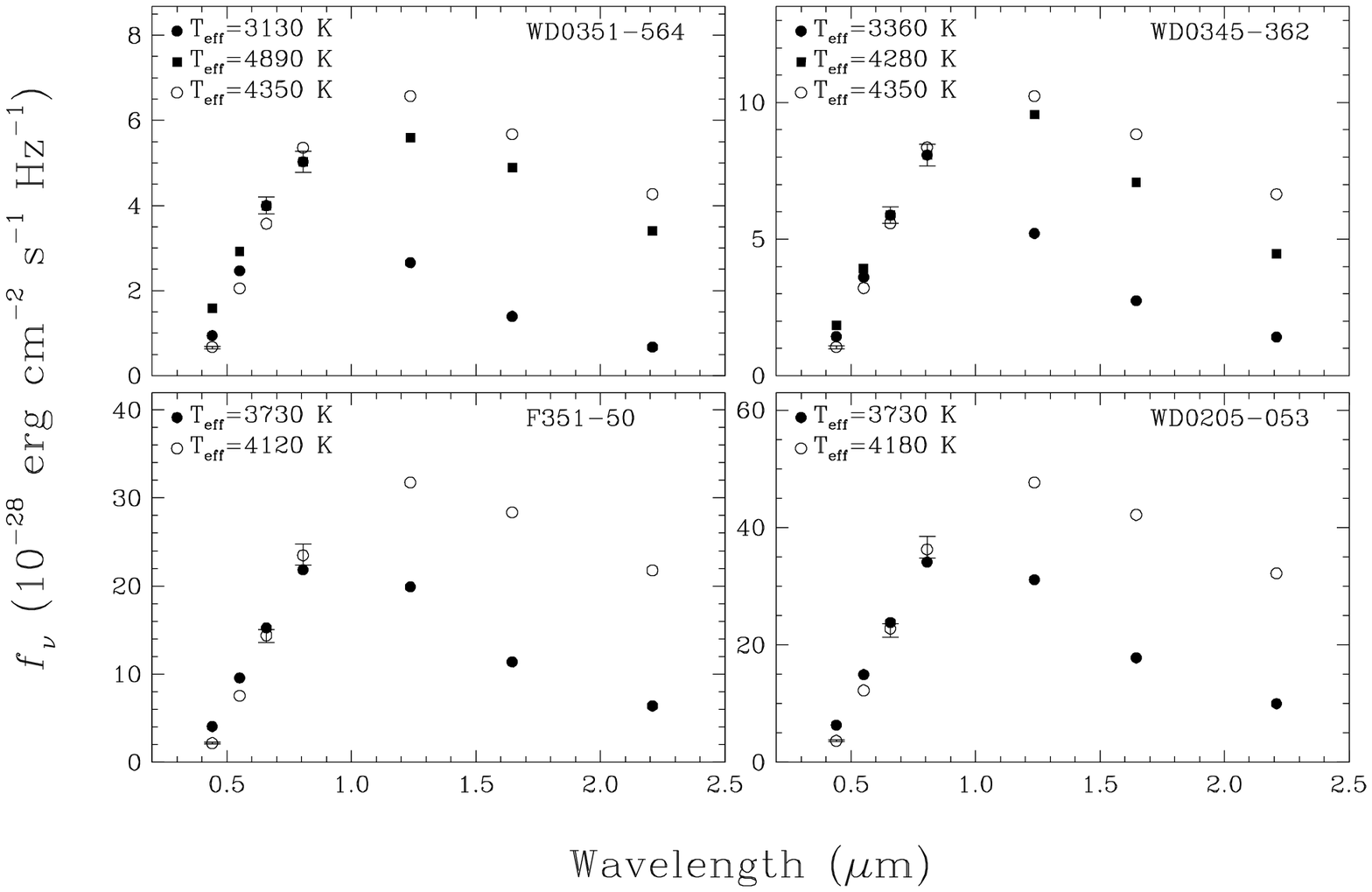] {Same as Figure \ref{fg:f7} but for the reddest 
objects shown in Figure \ref{fg:f5}. Note that the $B$ magnitude is
not used for the fits with pure hydrogen models ({\it filled
symbols}). Some objects have also two possible hydrogen
solutions.\label{fg:f8}}

\figcaption[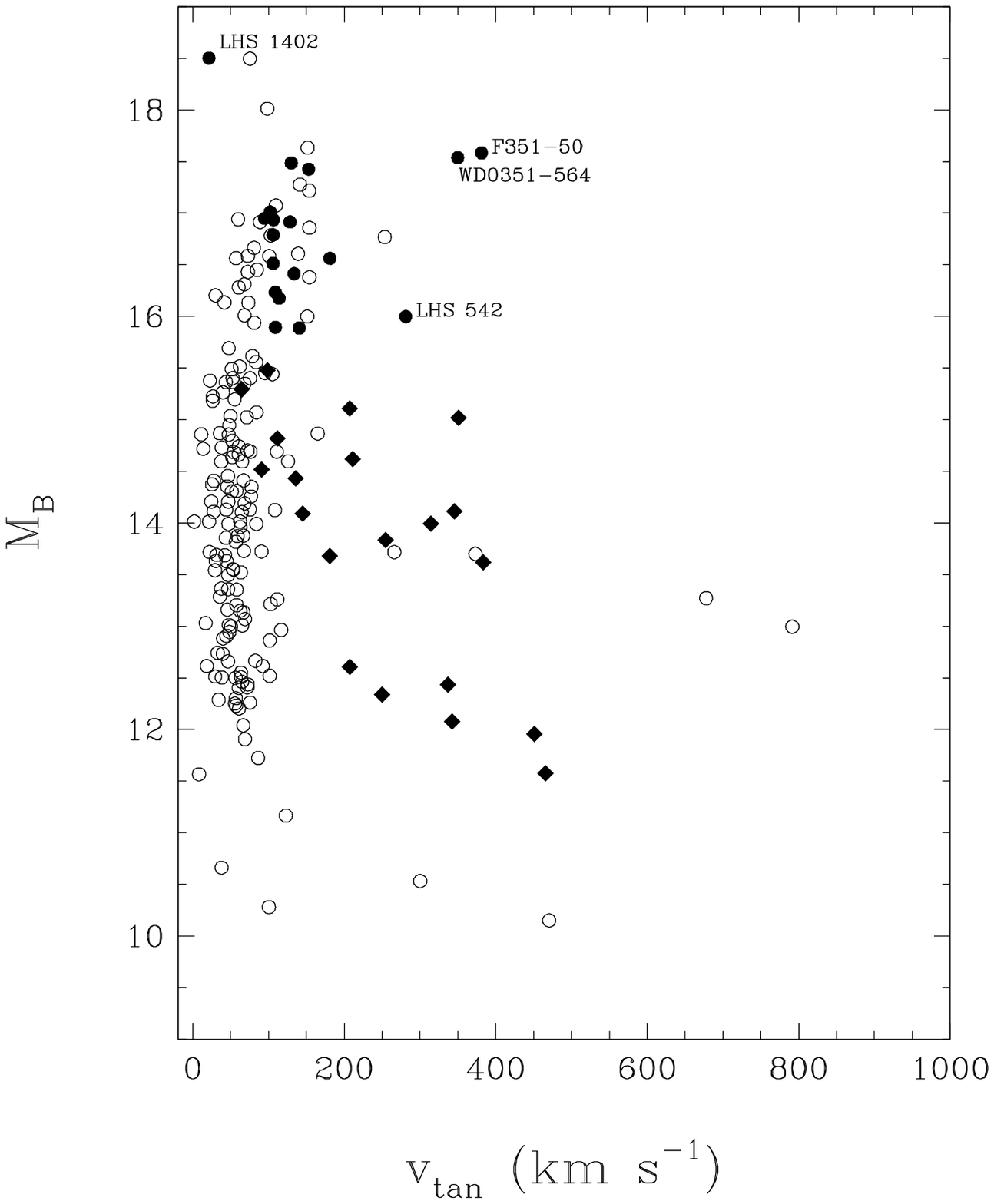] {Distribution of tangential velocities ($\vtan$) 
with the $B$ absolute magnitudes (\mb) for the white dwarfs of
OHDHS with $\Te>5000$~K ({\it filled diamonds}) and
$\Te<5000$~K ({\it filled circles}). The trigonometric parallax sample
of BLR shown in Figure \ref{fg:f1} is reproduced here as well ({\it
open circles}).\label{fg:f9}}

\clearpage
\begin{figure}[p]
\plotone{f1.eps}
\begin{flushright}
Figure \ref{fg:f1}
\end{flushright}
\end{figure}

\clearpage
\begin{figure}[p]
\plotone{f2.eps}
\begin{flushright}
Figure \ref{fg:f2}
\end{flushright}
\end{figure}

\clearpage
\begin{figure}[p]
\plotone{f3.eps}
\begin{flushright}
Figure \ref{fg:f3}
\end{flushright}
\end{figure}

\clearpage
\begin{figure}[p]
\plotone{f4.eps}
\begin{flushright}
Figure \ref{fg:f4}
\end{flushright}
\end{figure}

\clearpage
\begin{figure}[p]
\plotone{f5.eps}
\begin{flushright}
Figure \ref{fg:f5}
\end{flushright}
\end{figure}

\clearpage
\begin{figure}[p]
\plotone{f6.eps}
\begin{flushright}
Figure \ref{fg:f6}
\end{flushright}
\end{figure}

\clearpage
\begin{figure}[p]
\plotone{f7.eps}
\begin{flushright}
Figure \ref{fg:f7}
\end{flushright}
\end{figure}

\clearpage
\begin{figure}[p]
\plotone{f8.eps}
\begin{flushright}
Figure \ref{fg:f8}
\end{flushright}
\end{figure}

\clearpage
\begin{figure}[p]
\plotone{f9.eps}
\begin{flushright}
Figure \ref{fg:f9}
\end{flushright}
\end{figure}

\end{document}